# High-Resolution Near-Field Microwave Imaging using Loaded Circular Aperture Probe

Zubair Akhter, *Member, IEEE*, Mohammed Saif ur Rahman, *Member, IEEE*, and
Mohamed A. Abou-Khousa, *Senior Member, IEEE*

*Abstract*— Circular aperture probes have been successfully used for high-resolution near-field microwave imaging. It has been established that circular aperture probes could conceivably provide higher image fidelity compared to the rectangular aperture probes used conventionally for near-field imaging. In this paper, it is proposed to further enhance the near-field imaging resolution and sensitivity by loading the circular aperture with a resonant iris. The proposed probe herein operates in X-band frequency range and exhibits very localized near-field distribution at the opening of the iris. Consequently, its imaging resolution and sensitivity are enhanced compared to the conventional aperture probes operating over the same frequency band. The imaging capability of the proposed probe is analyzed using 3D electromagnetic simulation, and its performance is validated experimentally. The efficacy of the proposed probe for high-resolution imaging is demonstrated by imaging practical dielectric and metallic samples. Furthermore, the obtained images using the proposed probe are compared to those acquired using conventional circular and rectangular aperture probes. It will be demonstrated that the proposed probe provides higher sensitivity and resolution compared to the conventional aperture probes.

*Index Terms*— circular aperture probe, loaded aperture, near-field microwave imaging, resonant iris.

## I. Introduction

High-resolution near-field microwave imaging methods are emerging as powerful inspection tools for diverse industrial applications. These methods have been successfully applied for the nondestructive testing (NDT) of metal structures to map surface defects such as cracks [1]–[7], detecting precursor pits [8], [9], corrosion rust [10] under paint, and screening of composite materials for subsurface defects [11]–[13], among others. For these applications, the practical utility of near-field microwave imaging as an inspection modality and its merits compared to other well-established NDT modalities such as phased array ultrasonic testing have been demonstrated recently [11], [12].

The near-field microwave imaging probes developed in the past include coaxial probes [14]–[16], printed planar imaging probes [17]–[22], bulk periodic left-handed metamaterial (LHM) super lens [23], [24], non-periodic near-field plates [25], [26], waveguide-fed aperture probes [8]–[12], [27]–[29] and aperture probes loaded with resonant structures [2], [6], [30]–[33]. The near-field evanescent microstrip probes have been introduced as simple structures for high spatial resolution sub-surface imaging [20]–[22]. In these applications, the microstrip lines are tapered to cut off the fundamental propagating waves, to benefit from the high spatial resolution of the evanescent waves in the vicinity of the probes. The main design objectives of near-field imaging probes are to enhance the lateral resolution and sensitivity. Both parameters are usually improved by focusing the electromagnetic (EM) fields in the near-field region of the probe [23]. However, for the sub-surface imaging of dielectric material, increasing the depth of penetration inside the structure under inspection becomes essential as well. Therefore, it is imperative to maintain the focused probe footprint for a large range of imaging distances.

Waveguide-fed rectangular and circular apertures have been used extensively as a near-field imaging probe for many industrial applications [5], [11], [27]. These probes are typically devised for imaging as they are relatively inexpensive and easy to fabricate. Unlike most of other near-field probes, the near fields emitted from these apertures are confined to a small footprint in the near-field region and maintained as such for extended range of imaging distances. In essence, the footprint lateral size determines the resolution capability of the near-field probe. The obtained resolution from aperture probes is around half the largest dimension of the aperture [34]–[36]. The resolution of aperture probes can be further enhanced by loading the feeding waveguide with low-loss dielectric materials [6], [8], [37], [38], and tapering the feeding waveguide to yield a smaller aperture [8], [34], [39].

Recently, loading waveguide-fed rectangular apertures with resonant structures have been shown to significantly enhance the near-field imaging resolution of the probe [2], [6], [30]–[33], [40]. In [2] and [33], electrically small split-ring resonators were employed to load the rectangular apertures. The detection with such probes was performed via evanescent waves. These probes are very efficient for detecting surface defects such as metal cracks [2]. Since evanescent waves decay sharply as a function of the standoff distance, the detection of subsurface defects with such probes is still a challenge. On the other hand, the probe reported in [30] bids high spatial resolution for a wide range of imaging distances. However, the quality of the obtained images using folded-strip loaded aperture probe reported in [30] is compromised due to the multiple field peaks in the near-field distribution.

In this paper, a novel waveguide-fed circular aperture probe is introduced for high sensitivity and high-resolution near-field microwave imaging. The aperture of the proposed probe is loaded with a resonant iris which yields a very small footprint in the near-field region. The design of the proposed probe is relatively simple and provides higher sensitivity and resolution at larger standoff distances (SOD) compared to other waveguide-fed probes reported previously. Furthermore, unlike the previously reported loaded aperture probes, the proposed probe in this paper can be used for surface and subsurface imaging.

The remainder of the paper is organized as follows. Section II describes the design of the proposed probe along with a comprehensive characterization of its performance. Experimental probe characterization and imaging results are reported in Section III. Finally, Section IV concludes the paper.

## II. DESIGN & CHARACTERIZATION

### A. Probe Design

The proposed probe is basically a waveguide-fed circular aperture loaded with a resonant iris structure as shown Fig. 1. The proposed probe is simulated using 3D electromagnetic (EM) simulator, i.e., CST microwave studio [41], where feeding waveguide and iris are modeled as perfect electric conductor (PEC) in the initial investigation. The circular waveguide is completely filled with low loss dielectric material (i.e., Teflon) in order to lower the cut-off frequency of the dominant mode and isolate the probe from the subject under imaging. The feeding waveguide is designed with a diameter of $2a = 15.16$ mm to operate in the dominant mode (i.e., $TE_{11}$) with a cut-off frequency of around 8 GHz.

The circular aperture is loaded with an iris structure inspired from the literature [42]. The iris consists of a circular slot of radius $R$ and two metallic strips of width $W$ extended towards the center of the aperture as shown in Fig. 1(b). The gap between the extended strips at the center of the aperture is $g$. The circular slot with $R < a$ is primarily inductive. The capacitance required to resonate the iris structure is provided by the gap between the strips.

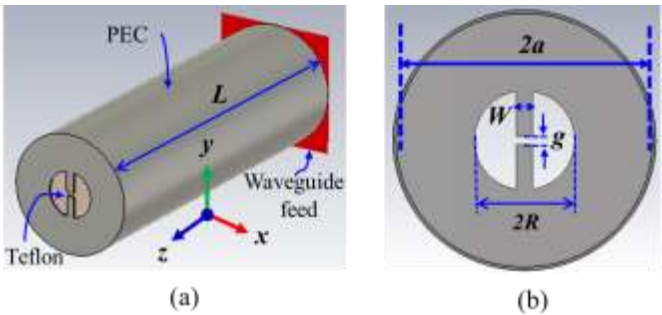

Fig. 1. 3D EM simulation model of the proposed probe (a) circular waveguide loaded with proposed resonant iris (b) detailed description of the resonant iris.

### B. Iris Design

The resonance frequency of the iris is a function of the circular slot radius $R$, strip width $W$, and the gap $g$ between the strips. At the resonance frequency, the magnitude of the reflection coefficient at the feeding port is minimized. The variation of the iris parameters and their effect on the resonance frequency of the probe is shown in Fig. 2. It is evident that the resonance frequency $f_r$ is inversely proportional to the slot radius $R$ and strip width $W$ while being proportional to the gap between the strip $g$. Increment in $R$ and $W$ increases the distributed inductance across the slot, while increment in $g$ decreases the distributed capacitance across the strips. As a matter of fact, the increment in distributed inductance or capacitance decreases the resonance frequency. Based on the parametric study, the final dimensions of the iris structure were chosen as $R = 2.975$ mm, $W = 1.0$ mm and $g = 0.5$ mm to yield resonance around 10 GHz (tabulated in Table I along with other pertinent model parameters). The magnitude of the reflection coefficient ($|S_{11}|$) for the simulated probe model with the selected set of iris dimensions is reported in Fig. 3.

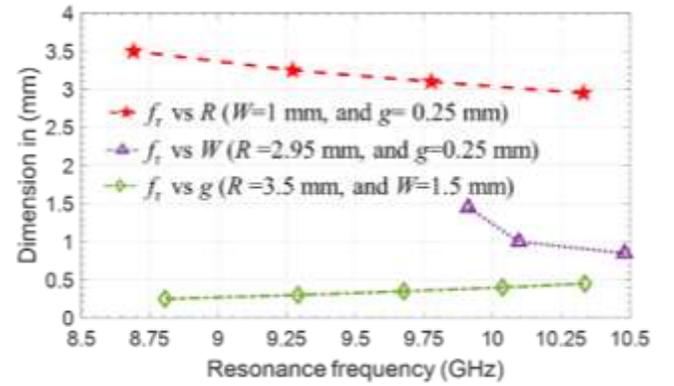

Fig. 2. Resonance frequency variation against iris dimensions ($R$, $W$, and $g$).

TABLE I
PROBE DESIGN PARAMETERS AND DESCRIPTION

| Symbol | Value (mm) | Description |
|---|---|---|
| $L$ | 40 | Length of the circular waveguide $L \approx 5\lambda_g/4$ at 10 GHz |
| $a$ | 7.58 | Radius of the circular waveguide |
| $R$ | 2.975 | Iris slot radius |
| $g$ | 0.5 | The gap between the strips |
| $W$ | 1 | Resonator strip width |
| $t_a$ | 0.2 | The thickness of the iris |

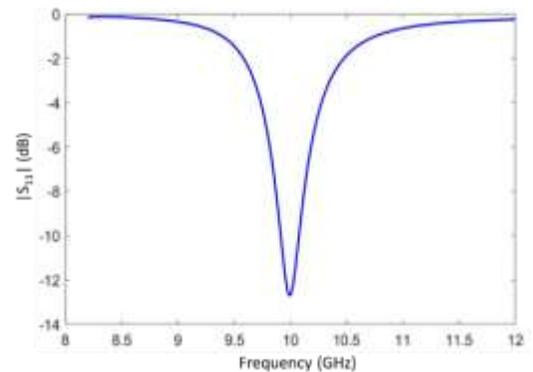

Fig. 3. Reflection coefficient magnitude (simulated) of the designed probe for $R = 2.975$ mm, $W = 1.0$ mm, and $g = 0.5$ mm.

*C. Near-field Distribution*

Near-field imaging probes are usually characterized based on their imaging footprint, which determines its ability to distinguish two adjacently placed targets and usually termed as lateral resolution [36]. This footprint can be estimated by observing the electric field magnitude in the near-field region at a given SOD from the aperture. Fig. 4 shows the spatial distribution of the electric field over *x-y* plane at SOD of 1 mm while the probe is radiating into free-space and operating at the resonance frequency. The electric field intensity in the gap region between the strips is very high and decays symmetrically in mutually orthogonal directions. Based on the footprint size in *x-y* plane, the designed probe is expected to provide lateral resolution of 1.8 mm along the *x*-axis and 1.5 mm along the *y*-axis (as per the standard 3 dB convention) at a standoff distance of 1 mm while operating at 10 GHz.

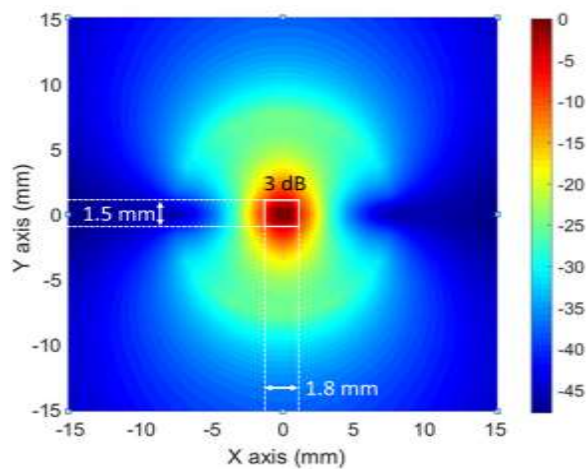

Fig. 4. Simulated electric field distribution (in dB) in the *x-y* plane at *z* =1 mm.

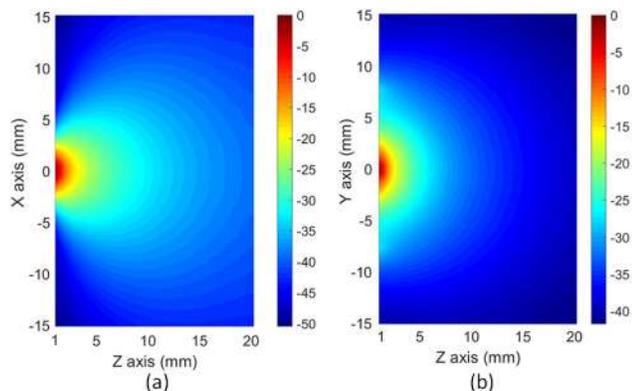

Fig. 5. Simulated electric field distribution (in dB) in the near-field of the probe in (a) x-z plane, (b) y-z plane.

The electric field distributions in the *y-z* and *x-z* planes are reported in Fig. 5(a) and Fig. 5(b), respectively. The electric field concentration near the iris structure and decay as a function of SOD is evident in these planes.

The sensitivity of the imaging probe to detect hidden targets placed away from the aperture depends on the magnitude of the electric field at target's location. In practical near-field imaging, the depth of investigation within the subject under imaging could extend to few tens of millimeters. In conventional aperture probe designs, there is a tradeoff between obtaining higher resolution (by making the aperture smaller) and sensitivity at large depth of investigation. The proposed probe design herein allows for tackling this practical tradeoff effectively. To demonstrate this, the relative capability of the proposed probe to irradiate targets at various distances from the aperture is compared to the conventional circular and rectangular aperture probes in terms of the electric field decay as a function of SOD. The considered circular aperture is of diameter 15.16 mm, and the rectangular aperture size is standard X-band (WR-90) waveguide aperture with dimensions 22.86 mm × 10.16 mm. Fig. 6 shows the normalized electric field magnitude as a function of the SOD for the three probes whereby the fields of these probes are normalized with respect to field at the center of the proposed probe. The electric field intensity at the center of the iris is around 21 dB higher than the field at the center of the circular and rectangular probes. The proposed probe provides higher field intensity in the close range up to 15 mm (0.5$\lambda$), and for farther ranges, yields similar field behavior as the other two probes. Despite the fact that the proposed probe has a smaller physical size than the other two probes, its sensitivity in the close range is expected to be higher than the other conventional aperture probes. Consequently, the proposed probe is expected to provide higher image dynamic range compare to the other probes especially for the near-surface defects.

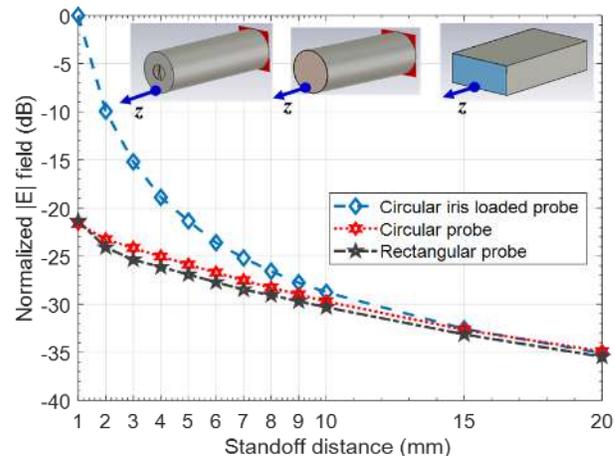

Fig. 6. Simulated E-field intensity as a function of standoff distance for the proposed probe and conventional rectangular and circular aperture probes.

To practically feed the designed probe with a 50 Ohm system, coaxial-to-circular waveguide transition was designed and incorporated into the simulation model. The feed was based on the 50 Ohm TEM to $TE_{11}$ transition reported in [43]. The two crucial parameters in such feed design are the coaxial insert length and position of the back-short [43]. As performed in [43], the feed design was optimized to yield a minimum reflection at the resonance frequency of the probe. In the proposed design, the coaxial insert length is kept at 5.08 mm while the position of the back-short is set to 8.22 mm.

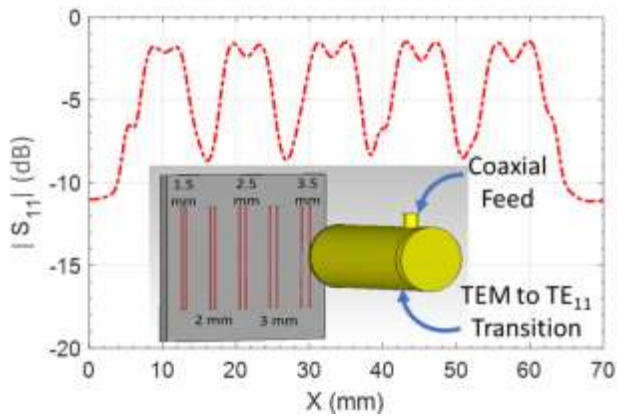

Fig. 7. Simulated line scan of wire targets with the simulation model in inset.

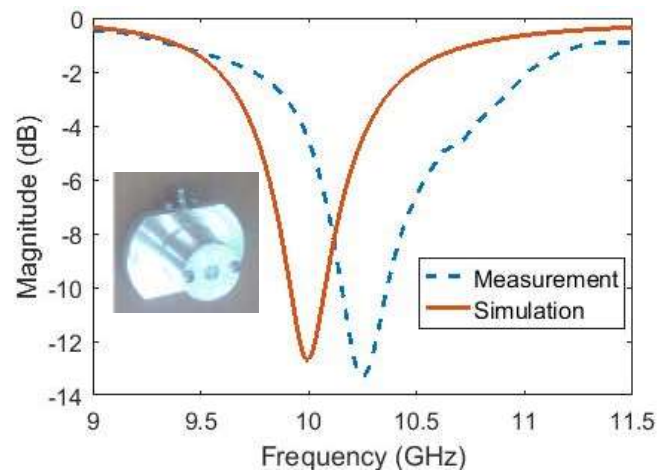

Fig. 8. Comparison of reflection coefficient (measurement and simulation) and probe photograph in the inset ($f_r$ =10.24 GHz).

After incorporating the feed, the resolution of the proposed probe was investigated by performing a line scan across thin conducting wires (0.4 mm diameter, and 40 mm length) with a step size of 0.25 mm at SOD of 1 mm. Five pairs of conducting wires are placed such that the distance between the wires (in pairs) is varying from 1.5 mm to 3.5 mm in steps of 0.5 mm as shown in the inset of Fig. 7. The probe was scanned over the wire pairs along the *x*-axis, and the magnitude of the reflection coefficient ($|S_{11}|$) was observed at each scan point. Fig. 7 shows the obtained line scan using the proposed probe. It is clear from the line scan that the designed probe could successfully distinguish the pair of wires with center-to-center spacing of 1.5 mm and larger, at SOD of 1 mm while operating at 10 GHz.

## III. MEASUREMENT RESULTS

### A. Probe Prototype

The designed probe with the parameters reported earlier was fabricated in house. The transition structure and feeding waveguide were fabricated out of aluminum block with the help of subtractive manufacturing process using a lathe machine. Electrical discharge machine (EDM) was used to fabricate the resonant iris. Mounting screws were used to attach the iris at the waveguide aperture to ensure proper alignment. At a later stage, the resonant iris structure was permanently attached to the circular waveguide aperture with the help of conductive epoxy. A photograph of the manufactured probe is shown in the inset of Fig. 8.

Prior to the imaging experiments, the reflection coefficient of the manufactured probe was measured using Keysight N5225A Performance Network Analyzer (PNA). Fig. 8 reports the comparison of measured and simulated reflection coefficient at the coaxial feed as a function of frequency. The overall shape of the measured response is similar to the simulated response. Due to fabrication tolerances, the resonance frequency is shifted slightly from the design frequency.

### B. Near-field Imaging Resolution

Line scans of eight pair of wires with varying interspacing from 1 mm to 8 mm in steps of 1 mm were acquired to determine the resolution of the proposed probe experimentally. These wire targets are 40 mm in length and 0.4 mm in diameter and embedded near the surface in a foam substrate. The wires were scanned using the proposed probe at SOD of 1 mm from the surface of the substrate and step size 0.25 mm. Fig. 9 and Fig. 10 show the wire pair arrangement and obtained reflection coefficient at each scan position. The probe was able to resolve wire targets interspaced by 2 mm, while it rendered just a single indication for the wire targets placed at 1 mm from each other. The measured results of the line scans are similar to those obtained in simulation. The resolution of the probe while operating at 10.24 GHz is around 2 mm ($\lambda/15$) as suggested in the simulation.

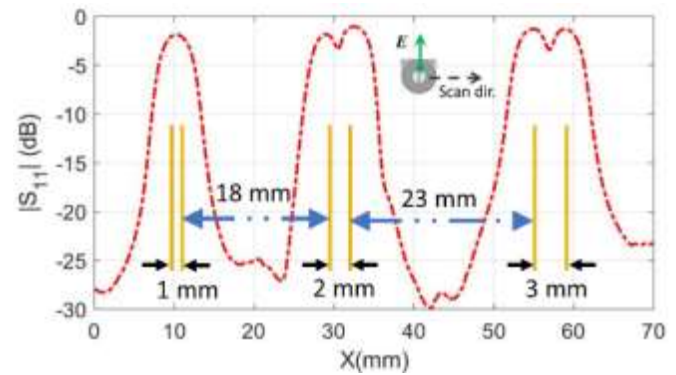

Fig. 9. Measured magnitude response ($S_{11}$) of the proposed probe for wire targets inter-spaced by 1 mm, 2 mm and 3 mm with iris orientation shown in the inset.

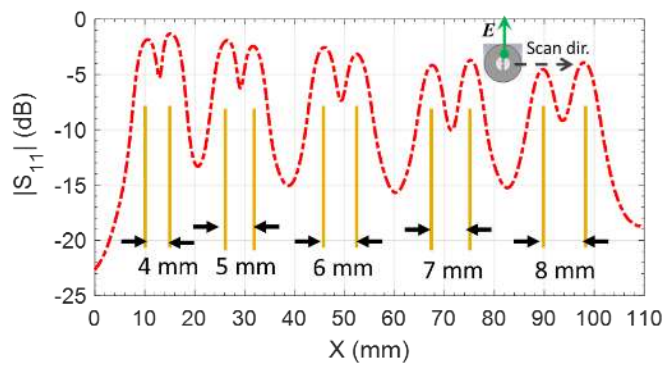

Fig. 10. Measured magnitude response ($S_{11}$) of the proposed probe for wire targets inter-spaced by 4 mm, 5 mm, 6 mm, 7 mm and 8 mm with iris orientation shown in the inset.

### C. Subsurface Imaging of Defects in Dielectric Sample

A dielectric sample of ABS material ($\varepsilon_r$ =2.5 and loss tangent $tan\delta$ = 0.0151 at 10 GHz [44]) was 3D printed using additive manufacturing process. Multiple flat bottom holes (depth 6 mm) were incorporated on the bottom side of the sample. The diameters of the holes range from 7 mm to 22 mm as reported in Fig. 11(a). The dielectric sample was scanned over a scan area 85 mm× 25 mm with step size 1 mm at a standoff distance of 1 mm from the top surface of the sample (i.e., the holes are hidden under 2 mm of ABS material as shown in Fig. 11(b)). Swept frequency reflection measurement over the range 8-12 GHz was carried out using Keysight N5225A PNA at each scan position. The images were formed at a single frequency point of interest whereby the phase and magnitude of the reflection coefficient were indexed in 2D map. For comparison purpose, the sample was scanned using the proposed probe, rectangular aperture probe, and circular aperture probe operating in the same frequency band. The measured aperture diameter of the manufactured circular probe is 15.30 mm while the measured rectangular aperture size is 22.70 mm × 10.10 mm. The rectangular aperture probe has a right-angled flange of 40.50 mm × 40.90 mm (end to end) with aperture located at its center. Figs. 12, 13, and 14 show the obtained (magnitude and phase) images using the proposed probe, rectangular aperture probe, and circular aperture probe, respectively.

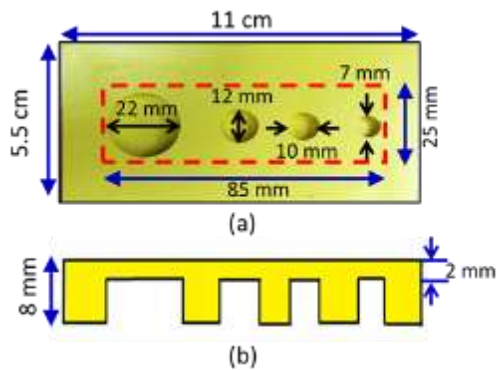

Fig. 11. 3D printed dielectric samples (a) bottom view (b) side cross-sectional view

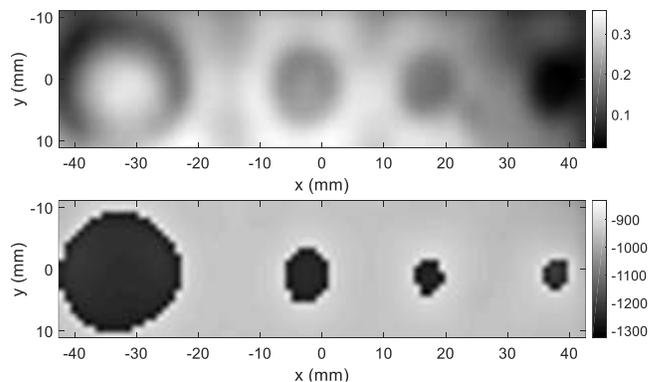

Fig. 12. Magnitude in dB (up) and unwrapped phase in degrees (down) images of the 3D printed sample using the proposed probe.

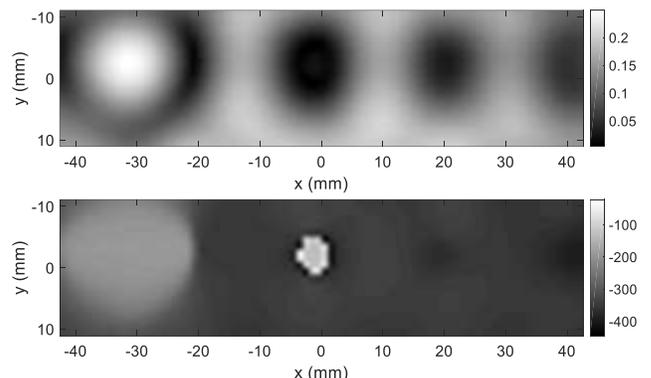

Fig. 13. Magnitude in dB (up) and unwrapped phase in degrees (down) images of the 3D printed sample produced by X-band rectangular aperture probe.

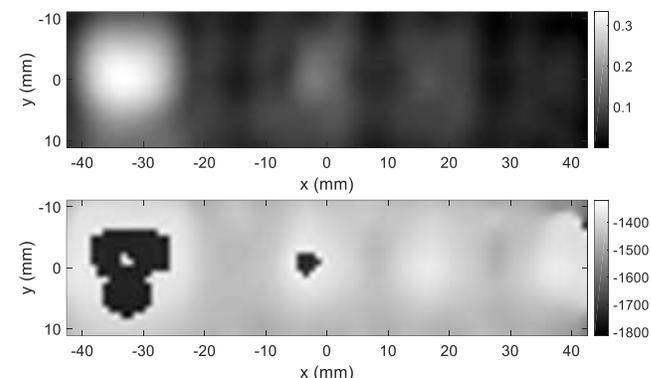

Fig. 14. Magnitude in dB (up) and unwrapped phase in degrees (down) images of the 3D printed sample produced by circular probe without iris.

It is evident that the proposed probe produced images of higher quality when compared to the other two probes. All defects were rendered clearly in the images obtained using the proposed probe with a significantly higher dynamic range than the other two probes (Figs. 13 and 14). Moreover, the shape of the defects is also accurately depicted in the image produced by the proposed probe.

### D. Corrosion Samples Imaging

Imaging surface defects in metals is considered very challenging, particularly behind a thin layer of corrosion and paint. To demonstrate the utility of the proposed probe for such imaging applications, two metal samples were considered (c.f. Fig. 15).

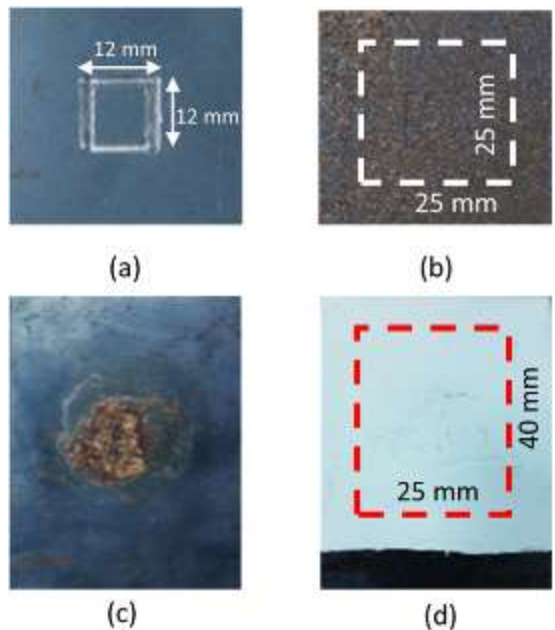

Fig. 15. Considered corrosion samples (a) sample #1 before corrosion (b) sample #1 after corrosion (with scan area highlighted) (c) sample #2 localized corrosion (d) sample #2 after painting the surface (with scan area highlighted).

In the first sample (sample #1), a square groove of lateral dimension 12 mm with 2 mm width and 0.5 mm depth was made. A photograph of the sample prior to corrosion is shown in Fig. 15(a). After scrapping the thin protective layer, the sample was exposed to the environment to corrode. Fig. 15(b) shows sample #1 after it corroded. In the second sample (sample #2), a limited area on the top surface of the metal was scrapped, and the sample was exposed to the environment. This resulted in the formation of a localized corrosion defect as shown in Fig. 15(c). Later, a paint layer of thickness 1 mm was applied on the specimen to conceal the corroded area (c.f. Fig. 15(d)).

The corrosion samples were scanned at a standoff distance of 1 mm around the defect area with 1 mm step size using the three probes as before. The first corrosion sample was scanned for an area 25 mm × 25 mm. The images produced using the proposed probe, rectangular aperture probe, and circular aperture probe are shown in Figs. 16, 17, and 18, respectively. The images shown in these figures are interpolated by a factor of 5 using the *imresize* function of MATLAB for better visualization. For the proposed probe, the resonance frequency at the first scan point is chosen for plotting the image, while the minimum reflection coefficient frequency is chosen at first scan point for the other two probes (rectangular and circular). It can be easily observed that the magnitude and phase images produced using the proposed probe provide a clear indication of the defected groove area despite the corrosion layer covering it. The other two probes provided faint or no indication of the groove defect in the presence of corrosion (c.f. Figs. 17 and 18).

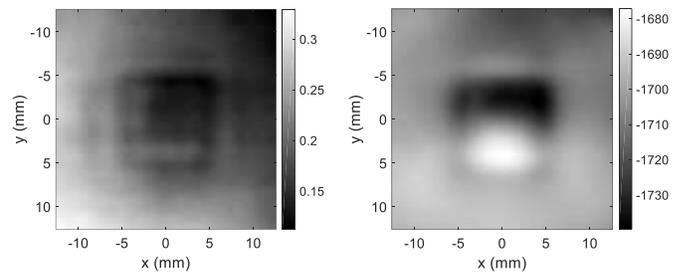

Fig. 16. Magnitude in dB (left) and unwrapped phase in degrees (right) images of corrosion sample #1 produced using the proposed probe.

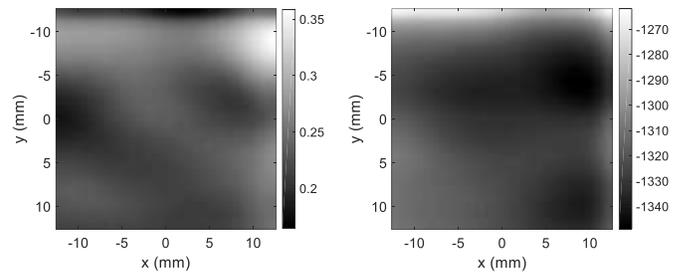

Fig. 17. Magnitude in dB (left) and unwrapped phase in degrees (right) images of corrosion sample #1 produced using X-band rectangular aperture probe.

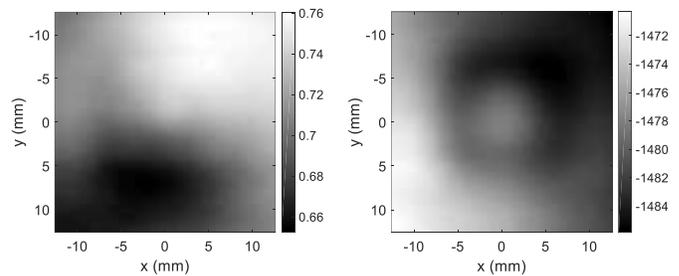

Fig. 18. Magnitude in dB (left) and unwrapped phase in degrees (right) images of corrosion sample #1 produced using circular probe without iris.

For the second metal sample, a relatively larger area was chosen for scanning (25 mm × 40 mm) to cover the defect covered with paint. The images produced with the proposed probe, rectangular aperture probe, and circular aperture probe are shown in Figs. 19, 20 and 21, respectively. The proposed probe produced a clear indication of the corroded area concealed behind the paint layer. On the other hand, the other two aperture probes showed faint or no indication of the corrosion under the paint layer (Figs. 20 and 21).

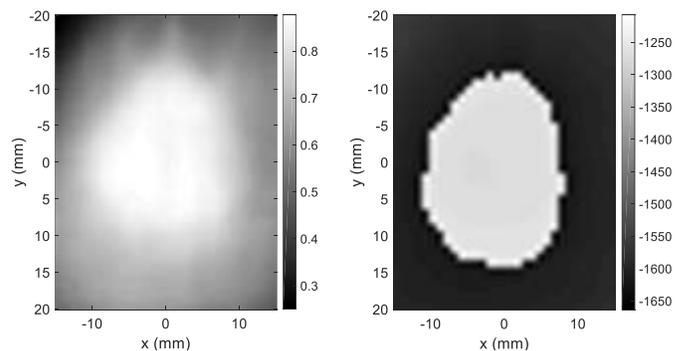

Fig. 19. Magnitude in dB (left) and unwrapped phase in degrees (right) images of corrosion sample #2 using the proposed probe.

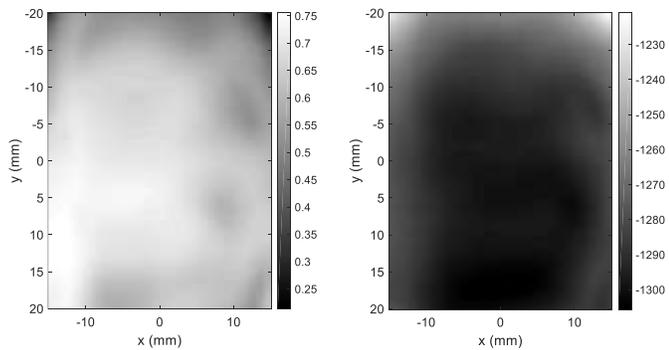

Fig. 20. Magnitude in dB (left) and unwrapped phase in degrees (right) images of corrosion sample #2 produced using X-band rectangular aperture probe.

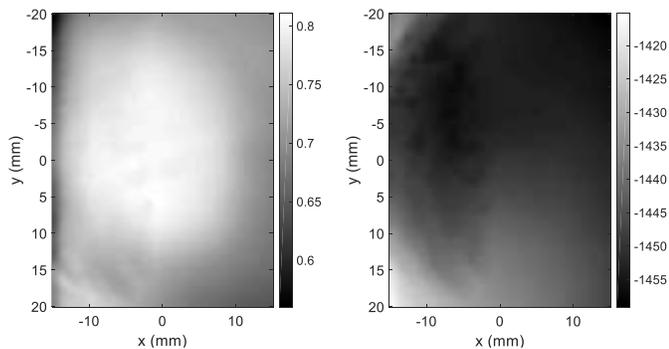

Fig. 21. Magnitude in dB (left) and unwrapped phase in degrees (right) images of corrosion sample #2 produced using circular probe without iris.

The comparison reported herein between the proposed probe, and two conventional aperture probes demonstrate the efficacy of the proposed probe for the imaging of subsurface defects in dielectric samples and near-surface defects in metallic samples. In fact, the images obtained using the proposed probe operating in the X-band frequency range (10.24 GHz) are comparable in quality to those obtained at much higher frequencies (Ka-band 33.5 GHz) using the aperture probes and phased array ultrasonic testing system [11].

IV. CONCLUSION

A new circular aperture probe loaded with resonant iris was proposed for near-field microwave imaging. The probe design was discussed, and its performance was thoroughly investigated using simulations and experiments. The near-field imaging performance of the proposed probe was demonstrated on metal and dielectric samples. The proposed probe offers high dynamic range and image resolution compared to conventional aperture probes such as the rectangular and circular aperture probes.